# Deformed states in paraelectric and ferroelectric nematic liquid crystals

Oleg D. Lavrentovich

Advanced Materials and Liquid Crystal Institute, Materials Science Graduate Program, Department of Physics, Kent State University, Kent OH 44240

Faculty of Chemistry, University of Warsaw, Zwirki i Wigury 101, 02-089 Warsaw, Poland

**Abstract**

Ground states of materials with orientational order ranging from solid ferromagnets and ferroelectrics to liquid crystals often contain spatially varying vector-like order parameter caused by inner factors such as the shape of building units or by the geometry of confinement. This review presents examples of how the shapes, chirality, and polarity of molecules and spatial confinement induce deformed equilibrium and polydomain states with parity breaking, splay, bend, and twist-bend deformations of the order parameter in paraelectric and ferroelectric nematic liquid crystals. Parity breaking results either from chirality of the constituent molecules, as a replacement of energetically costly splay and bend in paraelectric nematics, or in response to depolarization field in the ferroelectric nematic. Both paraelectric and ferroelectric nematics exhibit a splay cancellation effect, in which the elastic and electrostatic energies of splay along one direction are reduced by an additional splay along orthogonal directions.

## INTRODUCTION

A paraelectric uniaxial nematic (N) revolutionized the way we present information nowadays, enabling the industry of flat-panel and portable displays. The word "nematic" was derived by George Friedel from the Greek "$\nu\eta\mu\alpha$" which means a "tread" and refers to line defects often observed in unaligned N samples (1). The modern term for these defects is "disclinations". In the ground state of an N, the molecules, usually of elongated rod-like shape, point on average along the same direction, called the director $\hat{\mathbf{n}}$, **Figure 1a**. The director is a unit vector with the property $\hat{\mathbf{n}} \equiv -\hat{\mathbf{n}}$: even if the molecular heads and tails are different, their orientational order is insensitive to this distinction. There is no long-range positional order. Nematic molecules can carry dipole moments, but these are not required for the existence of N. Successful models of the N do not invoke the concept of permanent dipoles and operate with interactions such as steric repulsions and van der Waals attractions (2). Importantly, the N is formed by achiral molecules, i.e., each molecule is identical to its mirror image, or by the so-called racemic mixtures, with an equal number of left- and right-twisted enantiomers. Chiral molecules produce a twisted analog of N, called a cholesteric or a chiral nematic, N*, **Figure 1b**. Molecules of flexible dimeric and bent-core types yield a twist-bend nematic, $N_{TB}$, **Figure 1c-e** (3).

Because of the orientational order, the N exhibits anisotropy of properties such as dielectric permittivity and birefringence, which explains the success of the N-based liquid crystal displays (LCDs): An external electric field realigns the director within a micropixel of the LCD because of the dielectric anisotropy, while the birefringence changes the optical appearance of the pixel (4).

Gradients of the director field increase elastic energy. For deformations over length scales much larger than the molecular size, the elastic energy density $f_{FO}$ named after Frank and Oseen contains four terms: splay, twist, bend, and saddle-splay:

$$f_{FO} = \frac{1}{2}K_1(\operatorname{div}\hat{\mathbf{n}})^2 + \frac{1}{2}K_2(\hat{\mathbf{n}} \cdot \operatorname{curl}\hat{\mathbf{n}})^2 + \frac{1}{2}K_3(\hat{\mathbf{n}} \times \operatorname{curl}\hat{\mathbf{n}})^2 - K_{24}\operatorname{div}[\hat{\mathbf{n}}\operatorname{div}\hat{\mathbf{n}} + \hat{\mathbf{n}} \times \operatorname{curl}\hat{\mathbf{n}}] \quad (1)$$

with the elastic constants $K_1$, $K_2$, $K_3$, and $K_{24}$, respectively (5). The saddle-splay term can be decomposed into splay, twist, and a special term called a biaxial splay (6) or $\Delta$ deformation (7).

In 1916, Max Born proposed that the nematic molecules that carry strong dipole moments that point in the same direction, thus effectively predicting the ferroelectric nematic $N_F$ (8). However, the N state was soon found in a material formed by quadrupolar molecules with zero dipole moments (9). For many years, the predicted $N_F$ did not find its experimental proof; ferroelectric ordering has been found only in the so-called smectic C (SmC) phases formed either by chiral rod-like molecules or by achiral molecules with a strongly bent core, see the recent reviews (3, 10). The spontaneous polarization, parallel to the smectic layers, is relatively weak since the ferroelectric order of dipoles is driven by molecular packing. The ferroelectric SmC did not enjoy the same level of practical utility as the paraelectric N since the periodic density modulations hinder uniform alignment and limit electric field-induced reorientation modes.

Year 2017 saw a new beginning, when synthesis and characterization of liquid crystals with large longitudinal molecular dipoles, on the order of 10 D, led to discovery of the $N_F$ with a spontaneous electric polarization **P** parallel to the director $\hat{\mathbf{n}} \equiv -\hat{\mathbf{n}}$ (11-14). The first synthesized $N_F$ materials are 2.3',4',5'-tetrafluoro[1,1'-biphenyl]-4-yl 2.6-difluoro-4-(5-propyl-1,3-dioxan-2-yl) benzoate (DIO) (11) and 4-((4-nitrophenoxy)carbonyl)phenyl-2,4-dimethoxybenzoate (RM734) (12), **Figure 1f**. The polarization density of $N_F$ is high, $|\mathbf{P}| \sim (3-7) \times 10^{-2}$ C/m$^2$, on the order of values found in solid ferroelectrics, and is remarkably sensitive to an external electric field. For example, a polarization of a uniformly aligned planar $N_F$ can be twisted away from the "easy axis" imposed by surface anchoring by a weak in-plane electric field, on the order of $10^2$ V/m (14). A similar realignment of a conventional N requires fields 1,000 times stronger (2).

The high polarization density poses a question about a stability of an $N_F$ "monocrystal" with $\mathbf{P}(x,y,z) = const$. Imagine a rectangular parallelepiped inside which $\mathbf{P}(x,y,z) = P\{1,0,0\}$ is parallel to the top and bottom faces and to two side faces. The remaining two $yz$ side planes are pierced by the polarization lines which deposit electric charges of density $\sigma = P \sim (3-7) \times 10^{-2}$ C/m$^2$. These charges produce a strong depolarization field $|\mathbf{E}_{dep}| = \frac{|\mathbf{P}|}{\varepsilon_0 \varepsilon} \sim (3-7) \times 10^8$ V/m, where $\varepsilon_0$ is the electric constant and the relative dielectric permittivity $\varepsilon$ is on the order of 10. The estimate $\varepsilon \sim 10$ is reasonable since once the polarization **P** caused by permanent dipoles is separated from other material properties, the remaining permittivity should not be much different from the one measured in a paraelectric N (15-17). The strong depolarization field, antiparallel to **P**, makes it difficult for the material to sustain its uniform polarization. For comparison, a much weaker electric field $E \sim 10^7$ V/m induces a ferroelectric order in a deep isotropic phase of $N_F$ material at a temperature 30 °C above the clearing point (18).

The depolarization field can be reduced by confining an $N_F$ slab between two short-circuited electrodes or by providing a large quantity of free ions. However, if these are not available, a geometrical solution is possible, through the formation of a domain structure, as first pointed out for a similar problem in ferromagnets by Landau and Lifshits (19) and Kittel (20). A similar phenomenon of spatially varying **P** is expected in the $N_F$; however, the geometry of it might be very different from the Landau-Lifshits-Kittel domains since the $N_F$ fluids are not hindered by crystallographic axes. The guiding agents of polarization patterns in the $N_F$ are (i) the long-range electrostatic interactions, (ii) surface anchoring caused by anisotropic interactions at interfaces with ambient media, and (iii) bulk elasticity of the gradients in molecular orientations.

Among the latter, a special place belongs to splay, which in an $N_F$ produces a bound charge of bulk density $\rho = -\mathrm{div}\,\mathbf{P}$. The bound charges experience long-range Coulomb interactions that increase electrostatic energy. It is thus expected that the polarization patterns in the $N_F$ should follow two rules: (1) **P** should be tangential to interfaces with dielectric media to avoid surface bound charge and (2) divergence of polarization in the bulk should be reduced to a minimum, $\mathrm{div}\,\mathbf{P} \to 0$, to avoid bulk bound charge.

The elastic and electrostatic energy cost of splay can be reduced by the so-called splay cancellation. In Cartesian coordinates, $\mathrm{div}\,\mathbf{P} = \left(\frac{\partial P_x}{\partial x} + \frac{\partial P_y}{\partial y} + \frac{\partial P_z}{\partial z}\right)$. Thus the undesired consequences of one-directional splay, say, along the $x$-axis, $\frac{\partial P_x}{\partial x} \neq 0$, can be mitigated by an *additional* splay along two other spatial directions, provided that the sign of these is opposite to that of $\frac{\partial P_x}{\partial x}$, e.g., $\frac{\partial P_x}{\partial x}\frac{\partial P_y}{\partial y} < 0$. Similar considerations apply to $\mathrm{div}\,\hat{\mathbf{n}}$ and elastic energy of the N.

The polarization field should depend on boundary conditions, such as the shape of confining volume and the direction of "easy axes" of surface anchoring. Confinement and anchoring necessitate deformations. Besides the splay, one expects twist, bend, and saddle-splay. Twist is of especial interest since it relates to chirality. The term "chirality" has been introduced by Lord Kelvin in 1894 as the property of asymmetry: a geometrical figure is chiral if its image in a plane mirror cannot be brought to coincide with itself (21). Chirality is inherent in many physical objects; in materials science, chiral structures impart unique optical, plasmonic, and biochemical properties. The first successful LCD was based on the chiral director structure of the so-called twisted nematic pixels.

The goal of this review is to consider how the electrostatic interactions, surface anchoring, confinement geometry, orientational elasticity and an external electric field shape the polarization patterns in the $N_F$. It is instructive to contrast the results with similar effects in a paraelectric N, which are described first.

1. **DEFORMATIONS OF PARAELECTRIC NEMATICS**

The ground state of a paraelectric N can acquire deformations in a number of ways, which can be classified as (i) intrinsic, rooted in molecular properties, and (ii) extrinsic, caused by external factors such as geometry of confinement.

*1.1. Intrinsic causes: molecular shapes.*

Molecules with a carbon atom connected to four different neighbors are chiral and impose chiral intermolecular torques onto neighbors. Local interactions create a slight twist that propagates along one direction in a chiral nematic, $N^*$, also called a cholesteric, referring to the cholesterol base of the first observed liquid crystals (2, 22, 23), **Figure 1b**. The twist can also propagate along two local directions, forming blue phases (24). Interestingly, recent numerical simulations of flexible rods with length polydispersity predict the occurrence of a cholesteric N* even when the rods are not chiral and show no other broken particle symmetry (25).

The $N^*$ twist elastic energy density is $\frac{1}{2}K_2(\hat{\mathbf{n}} \cdot \text{curl } \hat{\mathbf{n}} - q)^2$, where $q = \pm \frac{2\pi}{\mathcal{P}}$, $\mathcal{P}$ is the pitch, and the sign is determined by the handedness of twist. This energy vanishes for a helicoidal twist, $\hat{\mathbf{n}} = \left[\cos\left(\frac{2\pi}{\mathcal{P}}z\right), \sin\left(\frac{2\pi}{\mathcal{P}}z\right), 0\right]$, **Figure 1b**. The Frank-Oseen twist term resembles the Hamiltonian for elastic distortions in chiral ferromagnets with a so-called Dzyaloshinskii-Moria interaction (26, 27).

Chiral interactions are typically much weaker than the interactions favoring parallel arrangements, thus $\mathcal{P}$ is much larger than the molecular size (hundreds of nanometers vs nanometers). One way to explain this is to consider rotations around the long axes of chiral molecules which are not exactly rod-like but rather plank-like (28). If the interactions between molecules were fully symmetric around the long axes, the chiral interactions would be washed out.

If the interactions are slightly biaxial, then the chirality settles in; the weakness of biaxiality explains why $\mathcal{P}$ is large.

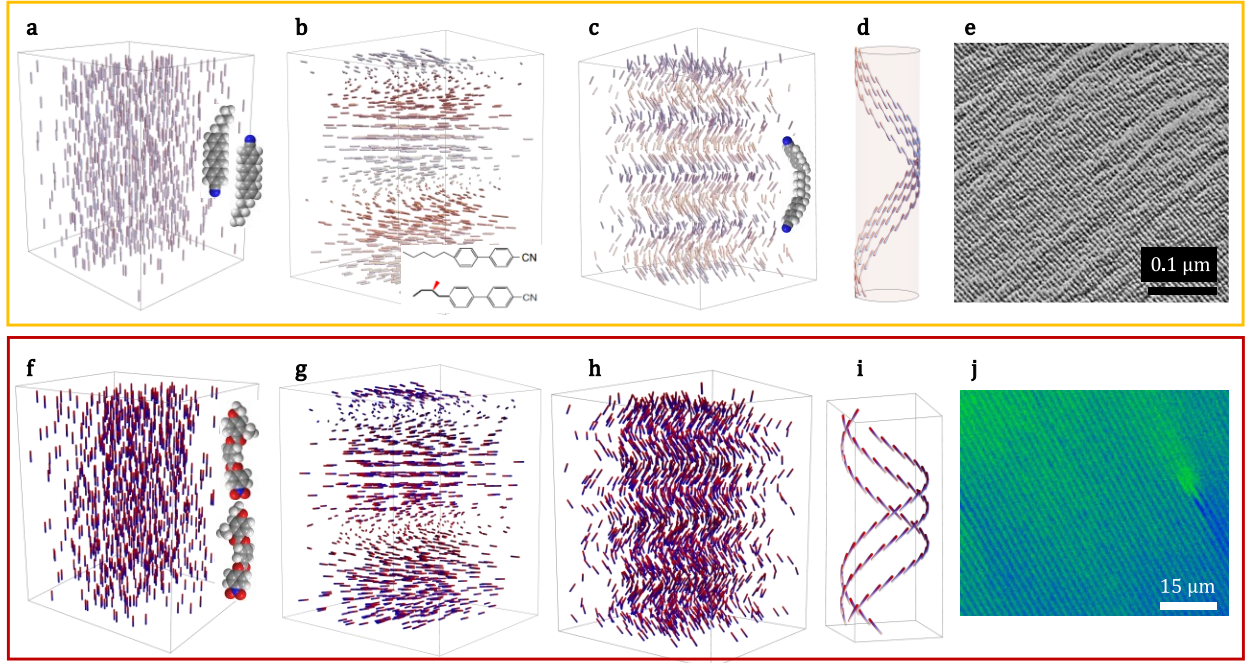

**Figure 1.** Nematics of (a-e) paraelectric N and (f-j) ferroelectric $N_F$ types. (a) Calamitic N formed by rod-like achiral molecules; the inset shows pentylcyanotriphenyl molecules. (b) Chiral nematic N* formed by chiral molecules or by a mixture of achiral and chiral molecules such as 4-cyano-4′-pentylbiphenyl (5CB) and (S)-4-cyano-4′-(2-methylbutyl)biphenyl (CB15), respectively, shown in the inset. (c) Twist-bend nematic $N_{TB}$ formed by flexible dimer molecules such as 1,7-bis-4-(4'-cyanobiphenyl)heptane (CB7CB) of a bent shape. (d) Constant bend of the director is supported by twist. (e) Electron microscopy texture of the $N_{TB}$ in CB7CB. (f) $N_F$ phase; the inset shows two RM734 molecules arranged in a polar fashion. (g) Chiral ferroelectric $N_F^*$ formed by doping material such as RM734 with chiral dopant such as CB15. (h) Twist-bend ferroelectric nematic $N_{TBF}$. (i) The $N_{TBF}$ structure is similar to that of $N_{TB}$ in (c) but the ordering of molecules is polar. (j) Polarizing microscopy texture of $N_{TBF}$ formed by 4'-(Difluoro(3,4,5-trifluorophenoxy)methyl)-2,3′,5′-trifluoro-[1,1'-biphenyl]-4-yl 2,6-difluoro-4-(5-propyl-1,3-dioxan-2-yl)benzoate (JK103) molecule. Part (e) adapted from Ref. (29). Panel e provided by Dr. M. Gao. JK203 was kindly provided by Dr. P. Kula.

If the molecules are not chiral but prefer bent conformations (3), a twist-bend nematic ($N_{TB}$) emerges, **Figure 1c,d,e**, predicted theoretically (30-33) and observed experimentally (29, 34, 35). The driving force is the tendency of $\hat{\mathbf{n}}$ to adopt a fixed curvature bend favored by the molecular shape; the twist is secondary, serving to maintain the same bend curvature in 3D space,

**Figure 1d**. The ensuing pitch is short, ~10 nm, i.e., a few molecular lengths (29, 35) since the twist-bend structure limits rotations of the bent molecules around their long axes (3). The $N_{TB}$ molecules are not chiral; thus, the spontaneous chiral symmetry breaking is ambidextrous, with samples split into left- and right-handed domains (29, 36, 37).

A structure similar to the $N_{TB}$ but with a much longer pitch and homogeneous chirality occurs when an $N^*$ material with a small $K_3$ is acted upon by an electric or magnetic field that tends to align the molecules along itself. The competition of this field alignment and the twist caused by molecular chirality is mediated by bend deformations; as predicted in 1968 by de Gennes (38) and R.B. Meyer (39), the balance yields a twist-bend structure in the certain range of the field amplitudes provided $K_3 < K_2$. The structure is called the oblique helicoidal cholesteric and abbreviated Ch$_{OH}$; it allows one to control selective reflection of light in a broad spectral range by an electric or magnetic field (40, 41).

The director in the N, $N^*$, $N_{TB}$, and Ch$_{OH}$ can be presented in a unified fashion as $\hat{\mathbf{n}} = \{\sin\theta\cos\varphi, \sin\theta\sin\varphi, \cos\theta\}$, where $\theta = 0$ in the N, $\theta = \pi/2$ in the $N^*$ and $0 < \theta < \pi/2$ in the $N_{TB}$; $\varphi = \frac{2\pi}{\mathcal{P}}z$. All these phases have their polar analogs in the ferroelectric nematic realm, **Figure 1f-j**.

### 1.2. *Extrinsic causes: confinement of N.*

Spherical N droplets dispersed in an isotropic fluid with tangential surface anchoring illustrate the effect of confinement on the equilibrium director most clearly. One possible structure is axially symmetric bipolar, with $\hat{\mathbf{n}}$ along the meridional lines, splay at the poles and bend at the equator. The splay and bend give rise to flexoelectric polarization, predicted by R.B. Meyer (42):

$$\mathbf{P}_f = e_1 \hat{\mathbf{n}} \operatorname{div} \hat{\mathbf{n}} - e_3 [\hat{\mathbf{n}} \times \operatorname{curl} \hat{\mathbf{n}}], \tag{2}$$

where $e_1$ and $e_3$ are flexoelectric coefficients of splay and bend, respectively (42),(43). Meyer supplemented his theoretical model with an observation of quadrupolar ordering of bipolar nematic droplets in an isotropic melt (42). Droplets in which the inner structure is of a dipolar symmetry form polar ferroelectric-like arrays (44).

Droplets with tangential anchoring often exhibit parity breaking revealed by optical activity, **Figure 2a**. Spontaneous twist replaces splay and bend since in most N materials, $K_2$ is noticeably smaller than $K_1$ and $K_3$. The director lines at the surface are along the loxodromes rather than the meridians, **Figure 2a-c**. Williams predicted that the director in the spherical droplet is twisted when $K_2 < K_1 - 0.431 K_3$ (45). The criterium was verified by exploring N droplets near the N-to-smectic A (SmA) transition, at which $K_2$ and $K_3$ diverge (46). Near the transition, the N droplets show a splay-bend structure with no twist, while at a higher temperature, where $K_2$ and $K_3$ diminish, the droplets experience a parity-breaking transformation. The twist angle decreases as one approaches the axis, **Figure 2c**. Similar parity breaking occurs in spindle-like N "tactoids" appearing as N nuclei in the isotropic phase (47-51). Spontaneous twists emerge also in N spheres with perpendicular anchoring (52-57), **Figure 2d**, and in cylindrical capillaries (58-62).

In 1974, Press and Arrott observed lens-shaped N droplets on a water surface (63); $\hat{\mathbf{n}}$ is tangential to the N-water interface and close to perpendicular at the N-air interface. These "hybrid" boundary conditions impose splay-bend in the vertical cross section. The lens' periphery yielded a radial $\hat{\mathbf{n}}$. A naively expected structure would be an axially symmetric $\hat{\mathbf{n}}$ with an in-plane radial defect of strength $m = +1$ at the center. The strength $m = +1$ means that as one circumnavigates the central core, the director projection onto the film's $xy$ plane, $\mathbf{n}_{xy} = \hat{\mathbf{n}} - \mathbf{v}(\hat{\mathbf{n}} \cdot \mathbf{v})$, turns by $2\pi$ in the same direction. Surprisingly, alongside the expected radial structures, more frequent were textures with a spiral $\mathbf{n}_{xy}$ (63). Numerical simulations revealed that the splay-bend is partially relieved by twist since $K_2 < K_1, K_3$. Press and Arrott also noticed a "splay cancellation" effect: an in-plane radial $\mathbf{n}_{xy}$ "produces a term in the divergence which offset the contribution to the divergence from the variation in the (director) from the top surface to the bottom surface" (63). In large-area hybrid aligned N films similar mechanisms yield periodic patterns of twisted stripe domains and square lattices of +1/-1 defects (44).

A different mechanism of spontaneous twist in confinement is illustrated in **Figure 2e,f**. An N slab is confined between two flat plates. The bottom plate sets an easy axis of alignment along the $y$-axis. In a flat sample, this alignment extends to the entire volume, **Figure 2e**; there is no deformation, $\hat{\mathbf{n}} = \{0,1,0\}$. If the plates form a wedge, the alignment that does not cause any deformation in the bulk is alignment along the $x$-axis, $\hat{\mathbf{n}} = \{1,0,0\}$. This alignment caused by nontrivial confinement can be called "geometrical anchoring" (64). The geometrical anchoring along the $x$-axis and the physical anchoring along the $y$-axis compete to reduce the elastic energy of splay and twist $F_{FO} = \frac{K_1}{2h}\gamma^2(1-\tau^2) + \frac{K_2}{2h}\tau^2$, calculated per unit area of the wedge of a local thickness $h$ assuming that the dihedral angle is small, $\gamma \ll 1$. If the structure is not twisted, $\tau = 0$, the elastic energy is that of splay, $F_{FO} = \frac{K_1}{2h}\gamma^2$. The occurrence of twist reduces the energy if $\frac{K_2}{K_1} < \gamma^2$. Similarly to the Williams criterium, this condition requires a small $K_2$. Twist effects caused by confinement are especially frequent in the so-called lyotropic chromonic liquid crystals (48, 50, 60, 65-70), in which $K_2$ can be one order of magnitude smaller than $K_1$, $K_3$ (71-73). Recently, a very small $K_2$ was measured also in surfactant-based lyotropic N with disk-like micelles (74). For more discussion of disparities of elastic constants and their effect on liquid crystal structures, see the recent reviews (75, 76).

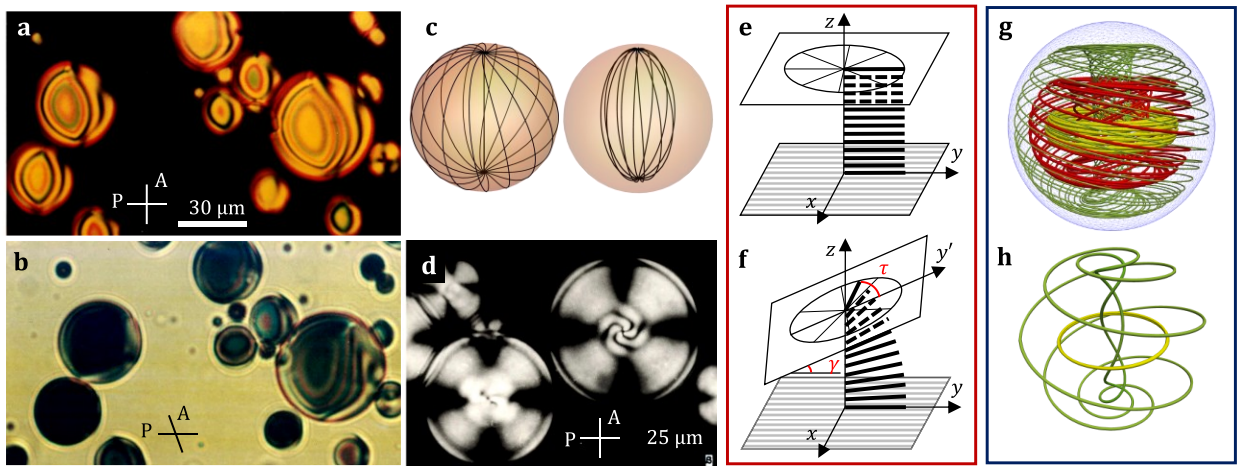

**Figure 2.** Twisted structures of spherical N droplets with (a,b,c) tangential anchoring and (d) homeotropic anchoring. (a) In observations with crossed polarizers P and A, the central part of the droplets is not extinguished but becomes so when (b) the polarizers are uncrossed, which indicates optical activity and thus the director's twist. (c) Scheme of the director at the surface and in the interior of a bipolar droplet. (d) Twisted structure in an N droplet with perpendicular anchoring at the interface. Materials: (a,b) n-butoxyphenyl ester of nonylhydrobenzoic acid in glycerol; (d), the same, with an addition of lecithin. (e,f) Mechanism of twist deformations caused by wedge confinement. In a flat sample (e) with one plate setting a planar orientation along the $y$-axis, the director is aligned uniformly along this axis. If the top plate with degenerate azimuthal alignment is tilted (f), the imposed splay is partially replaced with a twist by an angle $\tau$. (g) $N_F$ hopfion in a sphere with nested tori carrying dense sets of polarization lines; (h) polarization loops from tori in (g) are linked Part (d) adapted from Ref.(52), parts (e,f) adapted from Ref. (64), panels (g,h) reproduced from Ref. (77) [CC BY 4.0 International License] and kindly provided by I. Luk'yanchuk and A. Razumnaya.

## 2. DEFORMATIONS OF FERROELECTRIC NEMATICS

### 2.1. Intrinsic causes: molecular properties.

We first consider ferroelectric nematics with intrinsic deformations caused by molecular properties rather than confinement.

#### 2.1.1. Ferroelectric chiral nematic and twist bend ferroelectric nematic.

Chiral molecules added to the $N_F$ produce a ferroelectric chiral nematic $N_F^*$, **Figure 1g** (78-85). The true period of $N_F^*$ is $\mathcal{P}$ (rather than $\mathcal{P}/2$ as in $N^*$), as revealed by the network of dislocation in Grandjean-Cano wedges (79-81, 83, 84). The $N_F^*$ is much more sensitive to the electric field **E** applied perpendicularly to the helicoidal axis as compared to the $N^*$. Feng et al. (80) reported electrically tunable colors of the $N_F^*$ tunable by a field $(2-10) \times 10^4$ V/µm, which is ~100 times weaker than the field needed to modify the $N^*$ pitch. The difference is understandable since the realignment of paraelectric director is caused by the dielectric anisotropy coupling with the free energy density $\left[-\frac{1}{2}\Delta\varepsilon\varepsilon_0(\mathbf{E}\cdot\hat{\mathbf{n}})^2\right]$ quadratic in **E**, while the response of ferroelectric $N_F$ and $N_F^*$ is through the linear term $(-\mathbf{E}\cdot\mathbf{P})$; here $\Delta\varepsilon$ is the difference in dielectric permittivity measured along $\hat{\mathbf{n}}$ and perpendicularly to it and $\varepsilon_0$ is the electric constant.

The electric field applied perpendicularly to the helicoidal axis of either $N^*$ or $N_F^*$ inevitably adds higher-order structural modes (86) and thus diminishes the reflectivity. In this regard, $Ch_{OH}$ (87) and its polar analogs are preferable since the electric field acting along the helicoidal axis preserves the single mode structure and high reflectivity, which reaches 100% for obliquely incident light in a broad spectral range (41). There are two reports on synthesis of polar analogs of the $Ch_{OH}$, called a twist-bend ferroelectric nematic ($N_{TBF}$) by Karcz et al.(88), and a heliconical ferroelectric nematic abbreviated $^{HC}N_F$ by Nishikawa et al. (89), of the same oblique helicoidal supramolecular structure shown **in Figure 1h,i,j**. Unlike the $Ch_{OH}$, which is sustainable only in the presence of an electric or magnetic field, the newly discovered polar phases $N_{TBF}$ and $^{HC}N_F$ do not require external fields and are either enantiotropic (88) or monotropic (89). A similar oblique helicoidal structure but with a smectic layering perpendicular to the twist-bend axis has been discovered by Gigg et al.(90) and abbreviated $SmC_P^H$. In $N_{TBF}$, $^{HC}N_F$, and $SmC_P^H$, the pitch is in the order of hundreds of nanometers, which causes selective reflection of light (88-90). The long pitch demonstrates that the chirality of $N_{TBF}$ and $^{HC}N_F$ is rooted in polar interactions rather than in the bent shape of molecules. Indeed, the $N_{TBF}$ and $^{HC}N_F$ forming molecules are straight rod- or slightly plank-like with a strong longitudinal dipole, about 13 D (88, 89). When an electric field is applied, the pitch and the conic angle of $N_{TBF}$ and $^{HC}N_F$ both decrease till the polarization aligns fully along the field (88, 89). The operational voltages of selective reflection of light reported by Nishikawa et al. (89) are a few times smaller than those in the electrically tunable paraelectric $Ch_{OH}$ (41).

### 2.1.2. Antiferroelectric mesophases.

Exploration of mesogenic materials with large longitudinal dipoles uncovered proper ferroelectric smectic phases A, C, and chiral C, in which **P** is normal to the layers or tilted (88-98). These have been reviewed by Strachan et al. (91) and Liao et al. (99) and are not discussed here.

There is also an intriguing layered phase, labelled $M_2$ (11), $N_x$ (100), $SmZ_A$ (101), $N_S$ (102), and $M_{AF}$ (103) which often, if not always, separates the paraelectric N and ferroelectric $N_F$ on the temperature axis. It was first observed in DIO (11, 100, 104, 105), then in RM734 (106, 107), and other materials (90, 108-111). The abbreviations above reflect the perceived structure of this phase. A consensus is that this phase is antiferroelectric (100, 104, 105).

Resonant synchrotron-based small angle X-ray scattering (SAXS) demonstrated that the polarization pattern of the intermediate phase of DIO is periodic with a period $2w = 17.5$ nm (101). Within each period, there are two sublayers of opposite polarization, manifested by electron-density modulation in non-resonant SAXS. The density modulation explains the "smectic" assignment proposed by Chen et al. (101). Since the molecules are parallel to the layers, the phase is labeled $SmZ_A$ to contrast the name to SmA, where molecules are perpendicular to the layers (101); the subscript A stands for antiferroelectric. Nematics, including the $N_{TB}$ (29, 35), show no density variation.

The heads and tails of DIO and RM734 molecules are chemically different, resulting in pear-like molecular shapes, which can give rise to flexoelectric polarization in splay deformations. The possibility of periodic splay caused by flexoelectricity has been considered from the early days of $N_F$ studies (13, 102, 112-114), following the model of flexoelectric modulation of a paraelectric N proposed by J. Selinger et al. (33, 113, 115). In the $N_F$, the splay of **P** generates space charges $\rho = -\text{div } \mathbf{P}$. These space charges produce an internal electric field normal to the layers, **Figure 3a**, which acts to reduce the splay amplitude $\theta$ (101), **Figure 3b**. The efficiency of this suppression is determined by electrostatic and elastic energies. If the characteristic extension of splay along the normal to a polar sublayer is $\xi$, then the electrostatic energy per unit area of a sublayer is $\xi P^2/\varepsilon_0\varepsilon$, while the Frank-Oseen elastic energy of such a unit area is $\left(\frac{K}{\xi^2}\right)\xi \sim K/\xi$; minimization yields a characteristic polarization penetration length

$$\xi_P = \sqrt{\frac{\varepsilon_0 \varepsilon K}{P^2}} \qquad (3)$$

introduced initially in the studies of ferroelectric smectics (116). For a typical $K = 10$ pN, $\varepsilon = 10$, $P = 3 \times 10^{-2}$ C/m$^2$, one finds $\xi_P \sim 1$ nm. If $w \gg \xi_P$, the internal field is so strong that the polarization is along the layers practically everywhere, **Figure 3b**: electrostatics overwhelms the flexoelectric effect and yields the SmZ$_A$ (101). The electron density modulation is then caused by the variation of the magnitude of $|\mathbf{P}|$, **Figure 3b**. When $\xi_P \geq w$, which might be fulfilled when the polarization is weak, $\xi_P \propto 1/P$, one expects a periodic splay of $\mathbf{P}$ with polarity alternating from one sublayer to the next, **Figure 3a**. The subdomains of opposite polarity are separated by charged (positively in **Figure 3a**) $\pi$ domain walls.

The flexoelectric vs electrostatic balance considered by Chen et al. (101) is applicable to the experiments of Rupnik et al. (107) who studied RM734 doped with an ionic fluid 1-Butyl-3-methylimidazolium hexafluorophosphate (BMIM-PF$_6$) and of Ma et al. (117) who explored RM734 in cells with ionic polymers. In pure RM734, the intermediate phase exists only in a narrow temperature range, about 1℃ (106). The ionic fluid expands this range to tens of degrees. The ions could dramatically reduce the electrostatic effects and increase $\xi_P$. Thin films of a doped RM734 show periodic splay, the direction of which alternates from one sublayer to the next (107), **Figure 3a**. The period increases with the concentration $c$ of the ionic fluid and becomes optically detectable when $c = 5$ wt%. Polarizing microscopy observations suggest that the intermediate phase contains double splay along two directions rather than a single splay (107, 117). These observations are in line with the model of periodically modulated splay nematic proposed by Rosseto and Selinger that suggests that flexoelectric 2D splay is usually energetically preferable as compared to 1D splay (113).

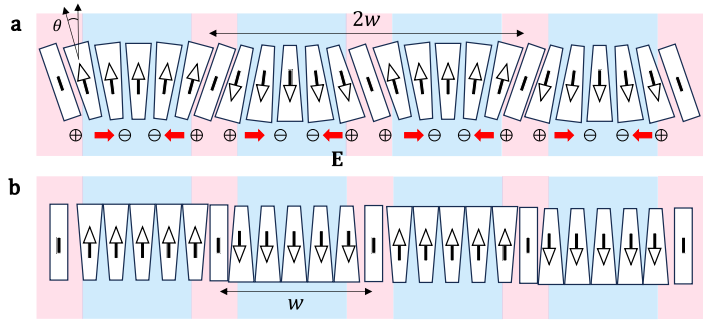

**Figure 3.** Periodically modulated antiferroelectric phases formed by asymmetric pear-like molecules that tend to splay periodically because of the flexoelectric effect. (a) Splay-induced bound charge and an intrinsic electric field along the layer that tends to reduce the amplitude $\theta$ of splay. (b) Proposed scheme of $SmZ_A$ (101). Adapted from Ref. (101).

In the double splay periodic structure, the splay polarity in two orthogonal planes is the same, $\frac{\partial P_x}{\partial x}\frac{\partial P_y}{\partial y} > 0$. In the splay-cancelling effect, the two splay polarities are of opposite signs, $\frac{\partial P_x}{\partial x}\frac{\partial P_y}{\partial y} < 0$. The double splay is favored by the flexoelectric effect, Eq.(2), while the splay cancelling is favored by electrostatics and by elasticity, as discussed later.

Experiments by Thapa et al. (84) on DIO doped with a chiral dopant show that the director in the chiral version of $SmZ_A$ twists along a single direction that is parallel to the layers' normal. The texture is thus compatible with the concept of 1D smectic-like periodicity along the direction perpendicular to smectic layers. Such a texture would be impossible if there were additional splay modulations within the layers: a twist of a periodic structure along any axis that is not normal to the layers violates equidistance of layers (2, 5). Dynamic light scattering experiments measured the compressibility modulus of the $SmZ_A$, $B = 4 \times 10^4 \frac{N}{m^2}$, which is about 100 times smaller than a similar modulus in SmA; the difference is understandable since the period of $SmZ_A$ is much larger than the period of SmA (118).

### 2.2. Extrinsic causes: confinement of $N_F$.
#### 2.2.1. Zenithal surface anchoring.

The zenithal part of anchoring requires **P** to be tangential to any interface with a dielectric medium. Consider a flat $N_F$ slab of a large $xy$ area. Imagine that **P** tilts away from the $xy$ plane by an angle $\psi$. The tilt would deposit bound charges $\pm P\sin\psi$ at the opposite sides of the slab and create a strong depolarization field $\frac{|\mathbf{P}|}{\varepsilon_0\varepsilon}\sin\psi \sim (10^8 \text{ V/m})\sin\psi$. This field acts to realign **P** back to a tangential state $\psi = 0$. In other words, the $N_F$ can cancel an electric field smaller than $\frac{|\mathbf{P}|}{\varepsilon_0\varepsilon}$ applied normally to an interface with a dielectric medium by a realignment of **P**; the effect is called "superscreening" (119). When the $N_F$ is confined in an S-shaped microchannel, superscreening forces **P** to follow the curved path remaining everywhere tangential to the lateral walls (119).

### 2.2.2. Azimuthal anchoring.

In-plane (azimuthal) anchoring of the $N_F$ depends on the adjacent medium. Monomolecular films of the $N_F$ preserve polar ordering when deposited onto an atomically smooth Au substrate (120). The common approach to align a liquid crystal, by rubbed polymer substrates, creates unipolar alignment of the $N_F$ (78, 121), in which the equilibrium orientation of **P** is either antiparallel to the rubbing direction (DIO at a polyimide PI2555 (122, 123)) or parallel to it (RM734 at PI2555 (124)). A planar cell formed by two unidirectionally rubbed polymer coatings assembled in an "antiparallel" fashion, produces a π- twisted **P** (14, 78). Unrubbed polymers such as polystyrene yield a degenerate azimuthal anchoring both in the N (125) and $N_F$ (126) materials. Below, we consider thin slabs of the $N_F$ with different types of azimuthal anchoring: fully degenerate, hybrid (one plate is degenerate, the second plate is planar), planar apolar (both plates are unidirectionally photoaligned so that the easy axis is of an apolar director-like type), and, finally, planar polar (both plates are unidirectionally treated to produce a polar easy axis).

### 2.2.3. Pancake and spherical $N_F$ droplets.

Some materials exhibit the $N_F$ phase on direct cooling from the isotropic phase (127-131), which allows one to study $N_F$ nuclei surrounded by an isotropic melt. When there is no treatment of the bounding plates, large droplets are of a pancake shape with a circular polarization, $\mathbf{P} = \{P_r, P_\varphi, P_z\} = P\{0, \pm 1, 0\}$ in cylindrical coordinates (128, 129). Such a structure is expected since the trajectories of **P** are closed, never piercing the interfaces nor creating space charge. The pancake droplets sometimes show bend accompanied by twist (130) and splay (131). Splay is very rare, since it induces space charge $\rho = -\text{div }\mathbf{P}$ and increases the electrostatic energy (132). The latter is difficult to calculate because of nonlocality of interactions; examples of such calculations are presented in later sections (133). A short-cut proposed originally for ferroelectric smectics is to renormalize the splay modulus (134-136),

$$K_1 = K_{1,0}(1 + \lambda_D^2/\xi_P^2), \tag{4}$$

where $K_{1,0}$ is the bare splay constant, of the same order as in the N, $\lambda_D = \sqrt{\frac{\varepsilon\varepsilon_0 k_B T}{ne^2}}$ is the Debye screening length that depends on the concentration $n$ of ions; $e$ is the elementary charge, $k_B T$ is the thermal energy, and $\xi_P = \sqrt{\frac{\varepsilon\varepsilon_0 K_{1,0}}{P^2}}$ is the polarization penetration length, Eq.(3). $K_1$ in the N$_F$ is expected to be much larger than $K_{1,0}$, $K_2$, and $K_3$. Experiments on domain structures in planar DIO cells suggest $K_1/K_3 > 4$ (122).

Luk'yanchuk et al. (77) proposed a twist-bend, or Hopfion, model of polarization field for a solid ferroelectric nanosphere, **Figure 2g-j**. Inside the sphere, the polarization lines form closed dense trajectories on nested tori. If the polarization is upward at the Hopfion axis, it bends and turns downward at the surface, remaining tangential to it. These two antiparallel directions of **P** are reconciled by ±π twists as one goes from one torus to the next in the equatorial plane. Since the polarization is tangential to the surface, it contains two singularities at the poles, necessitated by the Poincare theorem, where the magnitude of polarization is reduced. The Hopfion is a topologically stable soliton formed by a knotted three-dimensional continuous vector field with bend and twist (either left- or right-handed) that cannot be unknotted without cutting the structure.

So far, there is no confirmation of Hopfions in N$_F$ droplets; however, they have been observed in fluid chiral ferromagnets (137) and cholesterics (138); for the review of these and other topologically states in paraelectric chiral liquid crystals, see Refs. (139-141).

### 2.2.4. N$_F$ films of large area with degenerate azimuthal anchoring.

One can argue that the main reason for predominance of bend in pancake droplets (127-131) is tangential surface anchoring. Experiments with large-area azimuthally degenerate N$_F$ films demonstrate that the bend is heavily favored over splay because of electrostatics rather than lateral anchoring. These films can be created by spreading a film of N$_F$ over the surface of isotropic fluid such as glycerol (122, 126), confining the N$_F$ between two polymer coatings such as unrubbed polystyrene (126, 143), or preparing freely suspended N$_F$ films (144). The polarization **P** aligns tangentially to the film and forms multiple circular "flux-closure" vortices, which do not create bound charge, **Figure 4a-d**.

The circular vortices show mostly smooth textures, **Figure 4a,c**. A topologically similar radial "aster" created by a photopatterned radial easy axis at the bounding plates is dramatically different: Instead of following the surface-prescribed splay, either $\{P_r, P_\varphi, P_z\} = P\{1, 0, 0\}$, or $P\{-1, 0, 0\}$ in the entire sample, the texture splits into multiple pie slices with **P** pointing inward and outward of the center, **Figure 4e**; in thicker samples, twists along the normal to the sample also emerge (133).

A domain wall that separates polarization fields $\mathbf{P}_1$ and $\mathbf{P}_2$ and minimizes its bound charge $\sigma_b = (\mathbf{P}_1 - \mathbf{P}_2) \cdot \hat{\mathbf{v}}_1$ is shaped as a conic section, either a hyperbola or a parabola; here $\hat{\mathbf{v}}_1$ is the normal to the wall, **Figure 4b**. To avoid bound charges, the conic bisects the angle between two neighboring polarization fields $\mathbf{P}_1$ and $\mathbf{P}_2$: the normal components of $\mathbf{P}_1$ and $\mathbf{P}_2$ along $\hat{\mathbf{v}}_1$ are continuous, while the tangential components change signs crossing the wall. A domain wall between two circular domains is hyperbolic (126,143), **Figure 4a,b**. When a circular domain borders a mostly uniform and slightly bend domain, the eccentricity reduces to 1 and the wall's shape is close to a parabola (126).

At real domain walls, **P** should realign smoothly rather than forming singular cusps. This smooth realignment produces a narrow "polarization-stabilized kink" with electric double layers stabilizing its width (144). The tips of the domain walls are often decorated with -1/2 disclinations connected by a circular arc at which $\mathbf{P}_1$ and $\mathbf{P}_2$ are antiparallel, **Figure 4f**. These disclinations reduce the elastic energy of the wall when the angle between $\mathbf{P}_1$ and $\mathbf{P}_2$ approaches 180° (142). Furthermore, expulsion of splay is not absolute in polygonal textures as some splay develops at the border of vortices with opposite sense of polarization circulation, **Figure 4c,d**. The corresponding localized bound charge can be screened by free ions.

Hedlund et al. (143) noticed that the thickness of the film near the core of circular vortices is reduced and associated this thinning with the core energy of vortex which pulls the free surfaces closer to each other. Such a thinning is visible in **Figure 4c**, thanks to the variation of interference colors. This attraction and thinning can be opposed by the increased area of the $N_F$-air interface. When the local thickness of the film and the length of the defect decreases by $\delta h$, the elastic energy gain is $K\delta h$. Assuming that the thickness of film varies linearly from $h - \delta h$ to $h$ as one moves away from the vortex core, the surface energy increase is $\Sigma R^2 \left(\frac{\delta h}{R}\right)^2 \sim \Sigma (\delta h)^2$, where $\Sigma$ is the surface tension at the $N_F$-air interface. Therefore, whenever $\delta h > K/\Sigma$, which for $K = 10^{-11}$N and $\Sigma = 10^{-3}$ J/m$^2$ means $\delta h > 10\ nm$, the surface tension can stop the thinning and the gain in elastic energy caused by the cores of vortices.

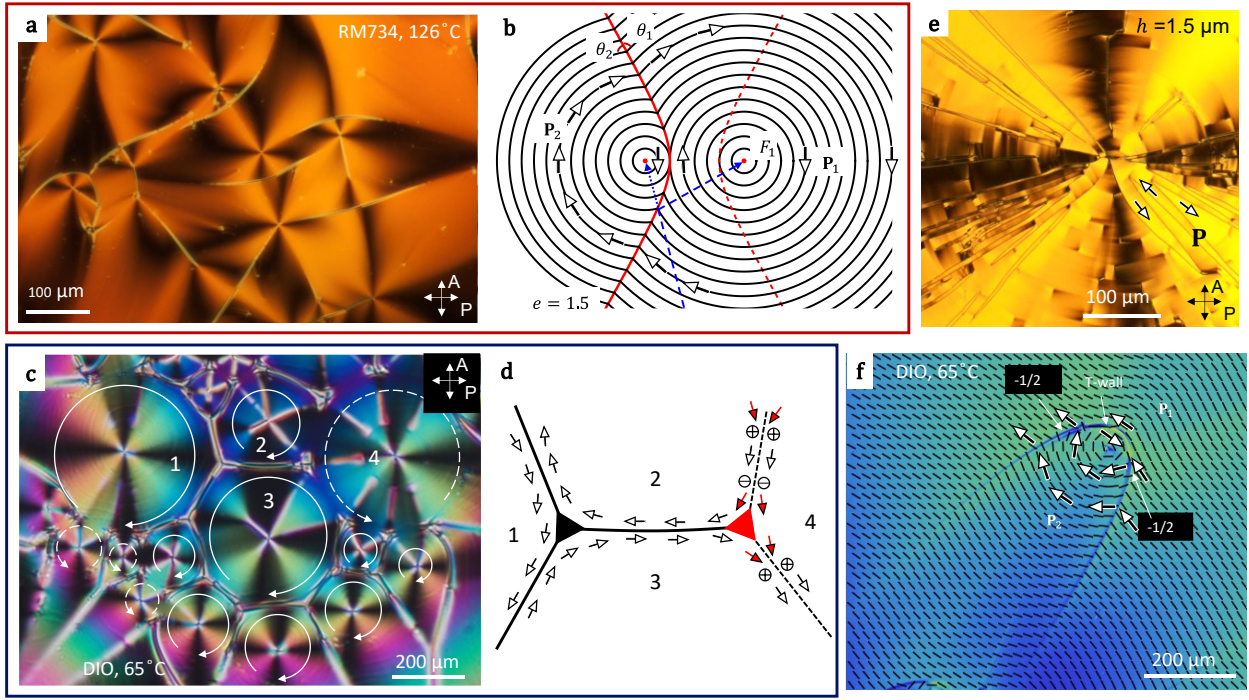

**Figure 4.** $N_F$ films with degenerate azimuthal anchoring. (a) Polarizing optical microscopy texture of circular vortices of topological charge $m = +1$; the polarization field $\mathbf{P} = \{P_r, P_\varphi, P_z\} = P\{0, \pm 1, 0\}$ bends; domain walls separating vortices are of hyperbolic shape. (b) Polarization field scheme of two circular vortices and a hyperbolic wall between them. (c) Polarizing microscopy texture of multiple $m = +1$ vortices; clockwise and counterclockwise circulations are shown by solid and dashed circular arrows. (d) Scheme of polarization pattern in vortices 1-4 in (c); red arrows of polarization exhibit splay and produce bound charges. (e) Radial "aster" induced by photopatterned radial "easy" axis at the bounding plates splits into multiple sectors with alternating polarity of $\mathbf{P}$. (f) A domain wall enclosing a circular vortex $m = +1$ is often decorated with two disclinations of strength $m_1 = m_2 = -1/2$ which relieve strong U-turn like bend of $\mathbf{P}$ near the tip of the domain wall. The disclinations are connected by a T-wall that flips $\mathbf{P}$ into $-\mathbf{P}$. Panels (a-d, f) adapted from Ref.(126), the micrograph in panel (e) provided by P. Kumari.

Conic sections are also frequent in smectic textures, as demonstrated by G. Friedel and F. Grandjean already in 1916 (145). Smectic layers are flexible but preserve equidistance. The director $\hat{\mathbf{n}}$ is normal to the layers and can experience splay but not twist nor bend, curl $\hat{\mathbf{n}} = 0$. The families of flexible equidistant surfaces form the so-called focal conic domains framed by focal lines shaped either as an ellipse-hyperbola pair or two parabolas (146). The conics thus emerge in the solenoidal field, div $\mathbf{P} = 0$, in the $N_F$, and in the irrotational field, curl $\hat{\mathbf{n}} = 0$, in the SmA.

### 2.2.5. $N_F$ cells with hybrid planar-degenerate azimuthal anchoring and spontaneous twist.

Imagine that the bottom surfaces of a large-area $N_F$ film imposes a unidirectional azimuthal anchoring on $\mathbf{P}$, while the second surface is tangentially-degenerate, allowing $\mathbf{P}$ to orient along any direction in the $xy$ plane. The unidirectional azimuthal anchoring makes it difficult to form circular domains. A uniform state is also unlikely because of the depolarization effects. Experiments demonstrate that the $N_F$ films split into a system of left- and right-twisted stripe domains, in which $\mathbf{P}$ twists along the normal to the film, **Figure 5a-d** (147). The twisted domains occur spontaneously as there is no surface-induced torque that facilitates their appearance. The spontaneous twists diminish when the $N_F$ is doped with the ionic fluid BMIM-PF$_6$ (147).

The twisted domains are caused by the material's tendency to reduce electrostatic energy. The electrostatic mechanism triggers chirality in a material formed by molecules with no chemically imposed chiral centers. Grönfors and Rudquist observed that the $N_F$ confined between two plates with obliquely evaporated SiO$_2$ layers also adopts spontaneous twists (148). The tendency is so strong that when an $N_F$ sample with twist domains is cooled into ferroelectric smectic phases, the polar smectic blocks preserve the inherited $N_F$ twist (103).

The reduction of $N_F$ electrostatic energy by twists was theoretically predicted by Khachaturyan in 1974 (149); the model was extended recently by Paik and J. Selinger (150) and by M. O. Lavrentovich et al. (133) who considered the screening effect of ions. The model considers an infinitely long circular cylinder and predicts a spontaneous twist of polarization, which rotates around the cylinder's $z$-axis, remaining perpendicular to it, $\mathbf{P} \propto [\cos(2\pi z/\mathcal{P}), \sin(2\pi z/\mathcal{P}), 0]$ (149). Although heuristically important as an illustration of the electrostatics-elasticity balance, in which electrostatics tends to reduce the pitch $\mathcal{P}$ and the elasticity tends to increase $\mathcal{P}$, the model is not applicable to experimental situations because it allows the polarization vector to pierce the cylindrical surface and thus to create surface bound charges.

### 2.2.6. N$_F$ cells with apolar planar anchoring

Theoretical and experimental analysis of the balance of long-range electrostatic, bulk elastic, and surface anchoring interactions was recently performed for DIO N$_F$ slabs confined between two plates with apolar in-plane anchoring (133). Such a film is created by photopatterning, using a thin film of an azobenzene dye that realigns along the direction perpendicular to the polarization of normally impingent UV light. When the polarization of light is uniform over the illuminated area of the empty cell with two azobenzene-coated plates, the induced "easy" directions are parallel to each other. These photoaligned apolar cells do not impose twist torques. When these cells are very thin, $h < 2$ µm, the textures split into uniaxially polar stripe domains; **P** is along the easy axis and reverses its polarity as one moves from one domain to the next, **Figure 5e,f,g**. There are no twists along the $z$-axis normal to the cell, except for a few regions marked by a letter "T" in **Figure 5f,g**. Thicker cells, $h > 2$ µm, show domains with a spontaneous twist along the $z$-axis, similar to the chiral domains in cells with hybrid unidirectional-degenerate anchoring, **Figure 5h,I,j** (147). The twist angle is close to 180°, as established by measuring the transmittance of light through the domains as a function of the angle between the polarizer and analyzer, **Figure 5k**.

The electrostatic energy is calculated as (133)

$$F_\rho = \frac{1}{8\pi\varepsilon\varepsilon_0} \iint \frac{\text{div}\mathbf{P}(\mathbf{r})\text{div}\mathbf{P}(\mathbf{r}')\exp(-\kappa|\mathbf{r}-\mathbf{r}'|)}{|\mathbf{r}-\mathbf{r}'|} d\mathbf{r}'d\mathbf{r} = \frac{1}{2\varepsilon\varepsilon_0} \int \frac{|\mathbf{k}\cdot\overline{\mathbf{P}}_\mathbf{k}|^2}{|\mathbf{k}|^2+\kappa^2} \frac{d\mathbf{k}}{(2\pi)^3}, \quad (5)$$

where the exponent describes screening effects by free ions, $\kappa = 1/\lambda_D \approx e\sqrt{n/\varepsilon\varepsilon_0 k_B T}$; the second equality presents the electrostatic free energy in Fourier space, $\overline{\mathbf{P}}_\mathbf{k} \equiv \int \exp(-i\mathbf{k}\cdot\mathbf{r})\,\mathbf{P}(\mathbf{r})\,d\mathbf{r}$. Electrostatic interactions prefer a spatial modulation of **P** with a small characteristic length scale $\xi$ (such as the pitch or domain's width). Conversely, any such modulation of **P** incurs an elastic or surface anchoring energy penalty that favors a larger $\xi$. The balance of electrostatic and elastic (or anchoring) energies generates a preferred $\xi$. There are two types of motifs in planar apolar cells (133): (a) domains of a uniform **P** along the apolar easy axis; **P** flips by $\pi$ when one moves from one domain to the next; (b) domains with left- and right-handed $\pi$-twists of **P** around the axis perpendicular to the cell's plane.

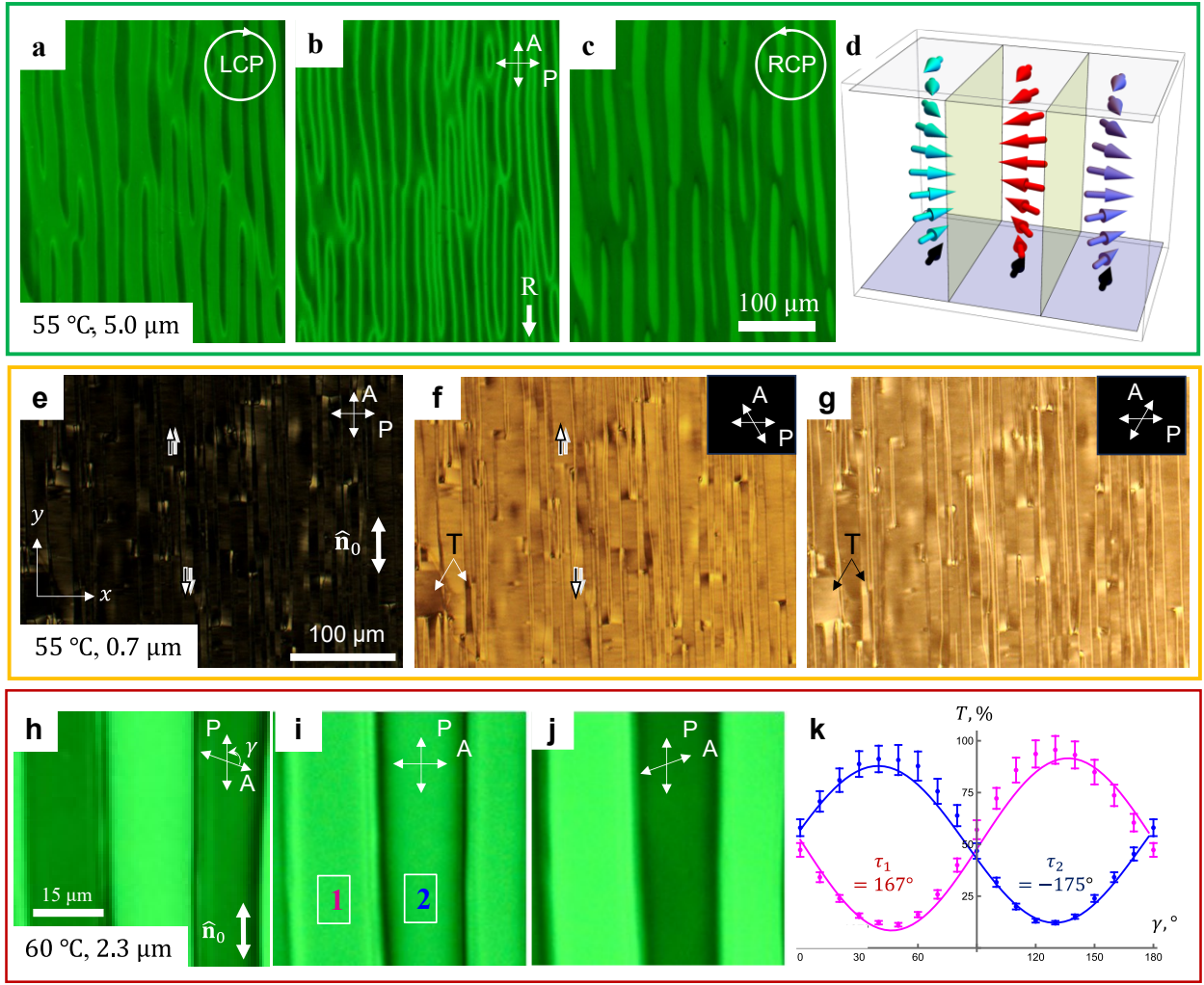

**Figure 5.** Domain structures of $N_F$ phase in DIO cells. (a,b,c) Chiral ground states of an $N_F$ slab confined between a unidirectionally rubbed polyimide PI2555 layer and azimuthally degenerate layer of polystyrene observed under an optical microscope with (a,c) circularly and (b) linearly polarized light. (d) Structure of domains with left-handed and right-handed $\pi$-twists. (e,f,g) Stripe domains with uniform polarization in submicron-thin photoaligned cells with planar apolar anchoring along the white double arrow $\hat{\mathbf{n}}_0$. Observations with (f,g) counterclockwise and clockwise uncrossing of analyzer and polarizer illustrate that **P** flips by $\pi$ from one domain to the next along the $x$-axis. (h,i,j) Chiral domains with $\pi$-twists of **P** along the $z$-axis normal to the field of view; observations with (h,j) decrossed and (i) crossed polarizers. (k) Transmitted intensity plot for domains 1 and 2 in (h,I,j) allows one to determine that the twist angle $\tau$ is close to $\pm\pi$. Panels (a-d) adapted from Ref. (147), (e-g) adapted from Ref.(133), panels (h-k) provided by P. Kumari.

As an example, the model (133) considers a single square domain of an area $L^2$ and a thickness $h$. If **P** is uniform, say, $\mathbf{P} = P(0,1,0)$, its electrostatic energy is $F_\rho = \frac{Lh\lambda_D P^2}{\varepsilon\varepsilon_0}$. However, if the polarization twists by $\pi$ between the top and bottom easy axes, $\mathbf{P} = P\left(\pm\sin\frac{\pi z}{h}, \cos\frac{\pi z}{h}, 0\right)$, the electrostatic energy reduces by a factor of 4: $F_\rho = \frac{Lh\lambda_D P^2}{4\varepsilon\varepsilon_0}$. The twisted structure also carries an elastic energy $F_{FO} = \frac{\pi^2 K_2 L^2}{4h}$. As the film thickness $h$ increases, the electrostatic energy increases $F_\rho \propto h$, while the elastic energy decreases, $F_{FO} \propto 1/h$. The balance yields a critical thickness

$$h^* = \frac{\pi}{P}\sqrt{\frac{2\varepsilon\varepsilon_0 K_2 Lh}{3\lambda_D}}. \tag{6}$$

Films thicker than $h^*$ are expected to be twisted. Even for a large domain, $L = 1$ mm, the critical thickness is small, $h^* = (1-10)$ µm, if one assumes $\lambda_D = (0.1-10)$ µm, $K_2 = 5$ pN, $P = 4.4 \times 10^{-2}$ C/m$^2$. This prediction agrees qualitatively with the experiments, in which the spontaneous twist appears in apolar planar $N_F$ cells thicker than a few micrometers, **Figure 5h,i,j** (133); see also Ref. (148) for spontaneous twists in polar planar cells.

The consideration above balances electrostatic and elastic energies for a single domain, which is assumed to be either uniform or $\pi$-twisted. It does not account for the possibility of spontaneous occurrence of multiple domains in ground states, similar to a lattice of left- and right-twisted domains in films with a degenerate anchoring at one surface (147). In polydomain textures, one must account for the presence of domain walls. The detailed internal structure of $N_F$ domain walls remains poorly explored. It should depend on the type of boundary conditions (polar, apolar, or degenerate), cell thickness, material parameters such as polarization, flexoelectric, and elastic constants, concentration of ions, and geometry of the domains. The wall might show complex 3D structure splitting into two surface disclinations (151) or forming a $2\pi$ twist disclination in the bulk (78). For example, in planar cells with antiparallel assembly, the $2\pi$ twist disclination wall running along the $y$-axis of surface anchoring and separating neighboring domains with a clockwise (CW), $\mathbf{P}_{CW} = P\left(\sin\frac{\pi z}{h}, \cos\frac{\pi z}{h}, 0\right)$, and counterclockwise (CCW), $\mathbf{P}_{CCW} = P\left(-\sin\frac{\pi z}{h}, \cos\frac{\pi z}{h}, 0\right)$, twists produces a strong surface bound charge $\pm 2P$ in the midplane $z = \frac{h}{2}$, where $\mathbf{P}_{CW} = P(+1,0,0)$ and $\mathbf{P}_{CCW} = P(-1,0,0)$ align head-to-head or tail-to tail, as in **Figure 5d**. This charge is reduced if the domain wall titls by $\pm 45°$ in the $xy$ plane of the cell, away from the $y$-axes (78), forming a zig-zag. Such a solution increases the length of the wall and its elastic energy. In similar experiments on planar polar cells with antiparallel assembly Sebastián et al.

(152) observed not only these zig-zag walls but also domain walls that run along the surface anchoring direction and perpendicular to it. In the case of hybrid polar-degenerate anchoring (147) and apolar planar anchoring (133), domain walls run along the easy axis; the reduction of space charge at the domain wall can be achieved either by tilting $\mathbf{P}_{CW}$ and $\mathbf{P}_{CCW}$ along the positive and negative directions of the z-axis normal to the cell (122) or through ionic screening.

Since the detailed domain wall structure is not known, difficulties remain in the theoretical description of the domain wall energy. If the wall splits into surface disclinations in thick samples, its energy per unit length might be independent of the cell thickness and scale approximately as an average Frank elastic constant, $f_{DW} \sim K$. Other models would imply some dependency of the domain wall on the cell thickness $h$, for example, $f_{DW} \propto \sqrt{h}$, when $\mathbf{P}$ realigns by $\pi$ through splay-bend across the wall (122) or as $f_{DW} \propto h$, when the wall structure does not change along the normal z to the cell; see Ref. (133) for more discussion.

### 2.2.7. Splay cancelling in planar $N_F$ cells under electric field.

The electro-optical applications of the N are based on the electrically controlled deformations of the director enabled by dielectric anisotropy; the relevant free energy density is $\left[-\frac{1}{2}\Delta\varepsilon\varepsilon_0(\mathbf{E}\cdot\hat{\mathbf{n}})^2\right]$, where $\Delta\varepsilon = \varepsilon_\parallel - \varepsilon_\perp$, $\varepsilon_\parallel$ and $\varepsilon_\perp$ are the permittivities measured parallel to $\hat{\mathbf{n}}$ and perpendicular to it, respectively. The realignment effect is called the Fréedericksz effect (2, 5). In the so-called splay Fréedericksz effect, a planar cell, $\hat{\mathbf{n}} = \{\pm 1,0,0\}$, is subject to an orthogonal electric field $\mathbf{E} = \{0,0,\pm E\}$, applied across the N slab sandwiched between two flat electrodes. If $\Delta\varepsilon > 0$, $\hat{\mathbf{n}}$ realigns towards the z-axis, when the field exceeds some threshold. The realignment is not sensitive to the polarity of the field. Immediately above the threshold, the deformation is that of splay. An intriguing question is what the response of an $N_F$ cell in a similar geometry would be. The response should involve a linear term $(-\mathbf{E}\cdot\mathbf{P})$ in the free energy density, which prevails at small fields. Splay of $\mathbf{P}$ might be difficult to create since it produces bound charge and a depolarization field that acts against the applied field.

Recent experiments (124) with a high frequency alternating current (ac) electric excitation of an $N_F$ planar cell, $\mathbf{P} = \{P,0,0\}$, show realignment of $\mathbf{P}$ with several unusual features, **Figure 6**. Namely, at any vanishingly small field, $\mathbf{P}$ oscillates with the frequency of the field, even if the latter is very high, hundreds of kHz (124). The time average orientation of **P** is still planar, i.e. the period-averaged tilt is zero.

When the voltage increases past some critical threshold, **P** acquires stationary periodic splay-twist deformations, **Figure 6a**. In the vertical $xz$ plane that contains **E** and the $x$-axis of surface alignment, splay is caused by tilt of **P** towards **E** in the bulk and by strong "electrostatic" zenithal surface anchoring that keeps **P** parallel to the electrodes at the surfaces. The stationary tilt of **P** alternates in space from "up" to "down" as one moves along the $y$-axis perpendicular to the rubbing direction, **Figure 6b**. The vertical splay develops mostly in the subsurface regions. In response to this field-induced splay, the system develops additional splay in the horizontal $xy$ planes, **Figure 6b**, in order to reduce space charge through the splay cancelling mechanism.

By the definition of divergence, the space charge depends on polarization variations along all three spatial directions,

$$\rho_b = -\text{div}\,\mathbf{P} = -\left(\frac{\partial P_x}{\partial x} + \frac{\partial P_y}{\partial y} + \frac{\partial P_z}{\partial z}\right). \tag{7}$$

Because of the vertically applied field, $\frac{\partial P_z}{\partial z} \neq 0$. Therefore, the space charge is reduced if an additional splay, $\frac{\partial P_y}{\partial y} \neq 0$, develops in the horizontal $xy$ plane, of a sign that guarantees $\frac{\partial P_y}{\partial y}\frac{\partial P_z}{\partial z} < 0$, **Figure 6b**. The horizontal splay alternates its polarity from V shape to Λ shape as one moves along the $y$-axis, so that the sign of $\frac{\partial P_y}{\partial y}$ is always opposite to the sign of the associated vertical splay $\frac{\partial P_z}{\partial z}$. The periodic up and down stationary tilts of **P** necessitate the appearance of left- and right-twisted domains that connect the splay-cancelling regions, **Figure 6b**.

The stationary tilt is induced by the balance of dielectric anisotropy and elastic torques (124) that is similar to the one in the paraelectric N, yielding $E_c = \frac{\pi}{h}\sqrt{\frac{K_{11}}{\varepsilon_0 \Delta\varepsilon}}$ as the threshold. The important difference is that the induced deformations in the N$_F$ are reshaped by the requirement to reduce the space charge, which makes the geometry spatially modulated rather than homogeneous as in the N.

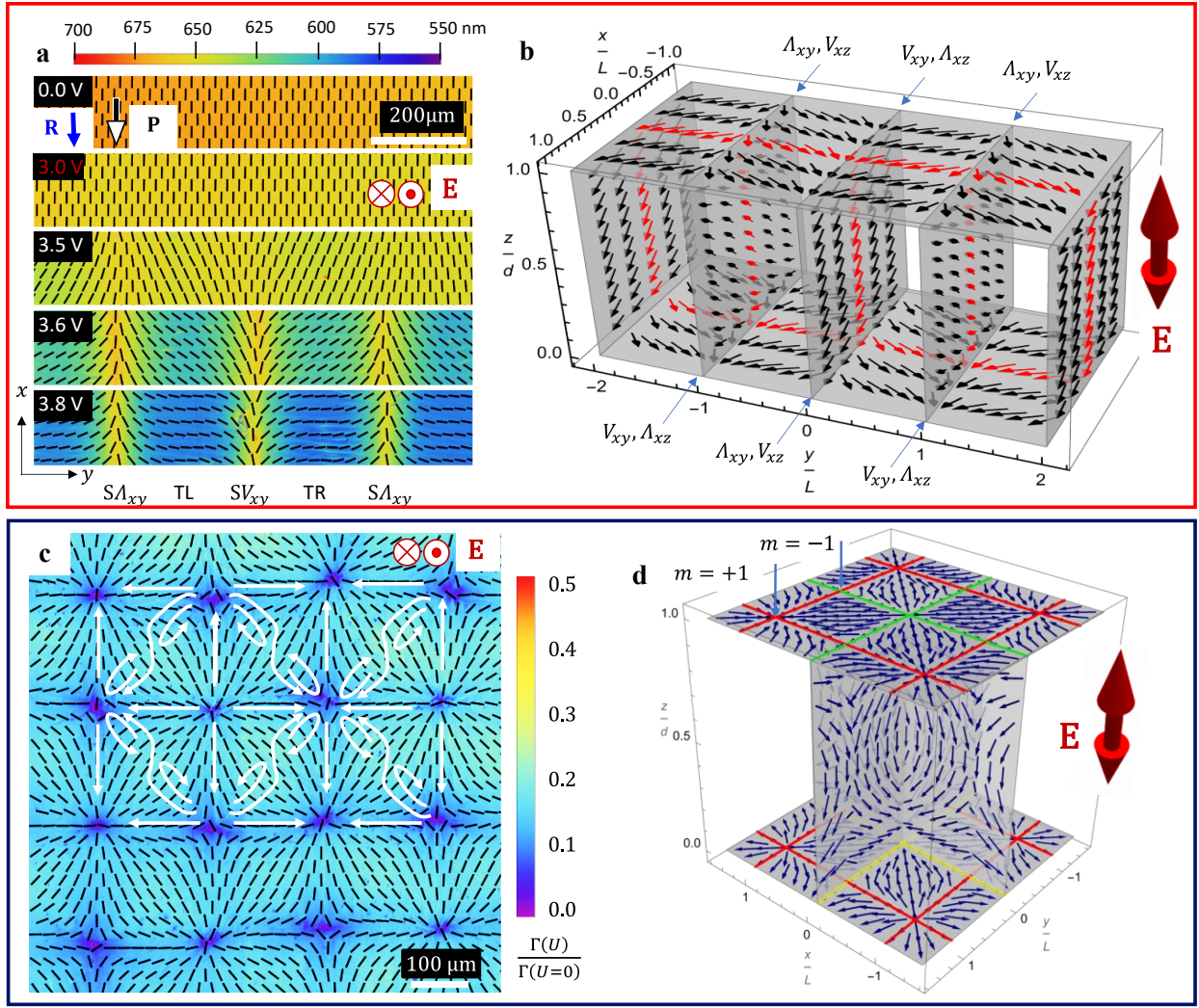

**Figure 6.** Periodic structures of splay Fréedericksz effect in the $N_F$ phase of RM734. (a) PolScope Microimager texture of a planar cell, $h = 3.5$ μm, that develops in-plane periodic splay above the voltage threshold of ~3.2 V; ac field of frequency 200 kHz. (b) Periodic splay cancellation: each splay region in the $xy$ plane is accompanied by a splay in the orthogonal $xz$ plane of opposite polarity. If the $xy$ splay appears as a letter V for a viewer, then the $xz$ splay resembles Λ, so that $\frac{\partial P_y}{\partial y}\frac{\partial P_z}{\partial z} < 0$ in all splay regions. The splay regions are separated by twists of left- and right handedness. (c) A high voltage 5.3 V, 200 kHz, produces a splay-bend pattern of +1/-1 defects with strong hydrodynamic flows sketched by white arrows, $h = 6.6$ μm. (d) A splay-bend lattice with negative bound charges in regions with green lines and positive bound charge in regions containing yellow lines; splay cancelling along the red lines. Microphotographs in panels (a, c) provided by B. Basnet, panels (b,d) adapted from Ref. (124).

Splay cancellation has been previously discussed for the director $\hat{n}$ fields in a hybrid aligned N (44, 63, 153) and suspended N (154) films and for periodic Fréedericksz transition in a polymer N with a large $K_1$ (155, 156). The pattern in **Figure 6a** with a large period, $2L \gg h$, is different from the patterns in a polymer N in which $2L \approx h$ (156). The large $2L/h$ can be caused by the large ratio of the azimuthal-to-zenithal deviations of **P** from the $x$-axis (124). Another difference is that in the N, the splay cancellation reduces the elastic energy density $\frac{1}{2}K_1(\text{div}\hat{\mathbf{n}})^2$, while in the $N_F$, it is an electrostatic effect. Electrostatic splay cancellation was suggested also for disclinations in a chiral $N_F^*$ (84), in which the externally imposed splay of a Grandjean-Cano wedge is relaxed by the in-plane splay of an opposite sign.

At still higher voltages, $U > U_{SB}$, where $U_{SB} \approx 2U_{ST} \approx 5$ V, **P** realigns fully along **E** in the middle of the cell and experiences splay and bend near the plates, forming a square lattice of +1 and -1 surface defects-boojums, **Figure 6c** (124). The stationary deformations are still accompanied by oscillations of **P**; the frequency range in which the square lattice forms is wide, from 10 kHz to 1 MHz. Similar splay-bend lattices of +1 and -1 defects with polarization oscillations have been observed in submicron cells with no surface alignment in the frequency range 20 Hz-2 kHz (157). Unlike the splay-twist stripes, the splay-bend lattice is accompanied by strong hydrodynamic flows, **Figure 6c** (124, 157). The splay cancellation is only partial, along red lines in **Figure 6d**. Regions around yellow and green lines in **Figure 6d** exhibit $\frac{\partial P_y}{\partial y}$ and $\frac{\partial P_z}{\partial z}$ of the same sign and produce positive and negative bound charges, $\rho_b = \pm P\left(\frac{\pi}{2\lambda} + \frac{1}{R}\right)$, respectively. The flexoelectric effect, which produces space charge $\rho_f \sim \frac{e_1}{\lambda^2}$ ($e_1$ is typically on the order of $10^{-11}$ C/m (5)), might facilitate the double splay when the deformation scale $\lambda$ is submicron. The positive and negative bound charges are facing each other across a small separation $h$, **Figure 6d**, which means that the applied ac field is rectified to produce a stationary potential difference at the electrodes. This potential difference of ~6-8V is an apparent reason for hydrodynamic flows that develop in the square lattices along closed loops (124). Oscillation of polarization is another factor facilitating the flows (157).

The splay-bend square lattice exhibits numerous intertwined mechanisms of coupling of the $N_F$ to an electric field: oscillations of polarization, dielectric-elastic balance that produces stationary splay, twist, and bend, accumulation of stationary potential differences at the electrodes as a result of deformations produced by a high-frequency ac voltage with zero dc component, potential flexoelectric effect, and, finally, occurrence of hydrodynamic flows.

*2.2.8.* A note on homotopy classification.

Homotopy groups approach has been used successfully to classify defects in various media, (158,159). It involves three steps (5). First, one defines the order parameter of the system and its degeneracy space, also called the order parameter space (OPS) $\mathcal{R}$, i.e. the manifold of all possible values of the order parameter that do not change thermodynamic potentials of the system. A spatially varying order parameter such as $\hat{\mathbf{n}}$ or $\mathbf{P}$ maps the points of real space into the OPS. Second, the mapping of interest in testing topological stability of a defect are those of $i$ - dimensional spheres enclosing the defect in real space. A point defect in a 2D system or a line defect in a 3D system is enclosed by a linear contour, $i=1$; a point defect in 3D can be enclosed by a sphere, $i=2$. Third, one defines the homotopy groups $\pi_i(\mathcal{R})$, the elements of which are mappings of $i$-dimensional spheres enclosing the defect in real space into the OPS. To classify the defects of dimensionality $t'$ in a $t$-dimensional medium, one needs to know the homotopy group $\pi_i(\mathcal{R})$ with $i = t - t' - 1$. Each nontrivial element of the homotopy group corresponds to a class of topologically stable defects; all these defects within the class are equivalent under continuous deformations of the order parameter. The elements of homotopy groups are topological invariants, which can be assigned topological "charges." The defect-free state corresponds to a unit element and zero topological charge.

In the $N_F$, the OPS in 2D is a unit circle, $\mathcal{R} = S^1$ and in 3D, a unit spere $S^2$. The point defects in 2D can be of any integer topological charge, positive or negative since $\pi_1(S^1) = Z$, where the elements are integers, corresponding to the integer topological charges $m$ of defects; half-integer charges are prohibited by the polar nature of $\mathbf{P}$. In 3D, any circumnavigated loop would have a contour in the OPS that can be shrunk into a point. It means that the $N_F$ does not support topologically stable disclinations, $\pi_1(S^2) = 0$. Thus, the term "nematic," introduced to stress the presence of disclinations, is not strictly applicable to the $N_F$. Of course, there might be vortices (142), stabilized by electrostatic rather than topological reasons, and surface disclinations stabilized by the surface anchoring (151). Their stability is energetical rather than topological. Topological point defects in the $N_F$ bulk, called hedgehogs, are topologically possible, $\pi_2(S^2) = Z$, but they would carry an enormous electrostatic energy. Also, surface point defects-boojums (160) are topologically stable, representing half-hedgehogs when $\mathbf{P}$ is tangential to an interface.

Besides isolated topological defects, $N_F$ can carry composite defects "inherited" from the N phase during the N-to-$N_F$ phase transition. For example, a pair of ½ disclinations can be connected by a wall defect in **P** (161), **Figure 4f**. Wall defects with a $\pi$-flip of **P** terminating at pairs of $\pm 1/2$ disclinations have been observed experimentally (142). E. I. Kats (162) classified the composite defects such as wall terminating at disclinations using the so-called relative homotopy groups (163) that allow one to predict which topological defects survive phase transitions such as the N-to-$N_F$ transition. Since the N contains topologically stable disclinations of half-integer charge, these defects can be preserved during the phase transition into the $N_F$ phase, but they would cease to be isolated (as $\pi_1(S^2) = 0$) and must be connected by domain walls in **P**. The homotopy classification also predicts the stability of Hopfions, **Figure 2j**, with integer topological charges, called the Hopf invariants, which are the elements of the third homotopy group, $\pi_3(S^2) = Z$ (162).

The homotopy classification of defects in the $N_F$ is severely limited by the electrostatic interactions which produce huge energy differences between defects that belong to the same topological class. For example, the circular vortex, written in the cylindrical coordinates $\mathbf{P} = \{P_r, P_\varphi, P_z\} = P\{0, \pm 1, 0\}$, and the radial defect, $\mathbf{P} = P\{\pm 1, 0, 0\}$, are formally of the same class with $m = 1$. However, the circular vortices carry mostly the elastic energy, while the radial ("aster") defects with the same $m = 1$ should carry a strong electric charge, concentrated at the core, $\rho(r) = -\text{div } \mathbf{P} = -\frac{P}{r}$. When imposed by photoalignment, these radial structures form "pie slice" domains in which the polarization alternates from $\mathbf{P} \parallel +\hat{\mathbf{r}}$ to $\mathbf{P} \parallel -\hat{\mathbf{r}}$, **Figure 4e** (133).

## CONCLUSIONS

Deformed equilibrium states of paraelectric and ferroelectric nematics form in response to a variety of intrinsic and extrinsic factors. The ferroelectric nematic shows a richer variety of deformed states than the paraelectric nematic. Among these are (i) domains in which the polarization is locally uniform but flips from **P** to $-\mathbf{P}$ from one domain to the next and (ii) domains with parity breaking in which **P** twists along an axis perpendicular to itself despite the absence of chemically chiral molecules. Splay in both types of nematics imposed by confinement or by external field can be replaced by twist, which is usually accompanied by a smaller elastic constant. A different mechanism mitigating the occurrence of splay is that of splay cancellation, which compensates for the divergence of a director or polarization along one direction with the divergence of an opposite polarity along orthogonal directions. The study of the deformed states in ferroelectric nematics is still in its infancy; for example, there is no experimental verification of the existence of Hopfions, **Figure 2g-j**. Further studies might result in discoveries of new modulated states, such as multiple twist grain boundary phases with polar interior, some of which have been reported very recently (164).

### DISCLOSURE STATEMENT

The author is not aware of any affiliations, memberships, funding, or financial holdings that might be perceived as affecting the objectivity of this review.

### ACKNOWLEDGEMENTS

The author is thankful to N. A. Clark, J. Gleeson, D. Golovaty, E. Górecka, A. Jákli, P. Kula, M.O. Lavrentovich, I. Luk'yanchuk, Yu. A. Nastishin, V.G. Nazarenko, S. Paladugu, S. Paul, D. Pociecha, S.V. Shiyanovskii, S. Sprunt, P.J. Sternberg, H. Wang, for useful discussions, to the graduate students B. Basnet, P. Kumari, K. Thapa, and M. Rajabi, who performed most of the experiments reviewed in this paper. JK203 was kindly provided by Dr. P. Kula. Part of this review was written during the author's sabbatical leave at the University of Warsaw; hospitality of E. Górecka and D. Pociecha is highly appreciated. The work was supported by NSF grant DMR-2341830.

### LITERATURE CITED

1. Friedel G. 1922. *Ann. Phys. (Paris)* 19: 273-474
2. de Gennes PG, Prost J. 1993. *The Physics of Liquid Crystals*. Oxford: Clarendon Press. 598 pp.
3. Jákli A, Lavrentovich OD, Selinger JV. 2018. *Reviews of Modern Physics* 90: 045004
4. Yang D-K, Wu S-T. 2006. *Fundamentals of Liquid Crystal Devices*. Chichester, England: John Wiley & Sons. 394 pp.
5. Kleman M, Lavrentovich OD. 2003. *Soft Matter Physics: An Introduction*. New York: Springer. 638 pp.
6. Selinger JV. 2018. *Liquid Crystals Reviews* 6: 129-42
7. Machon T, Alexander GP. 2016. *Physical Review X* 6: 011033
8. Born M. 1916. *Sitzungsberichte Der Koniglich Preussischen Akademie Der Wissenschaften*: 614-50
9. Sluckin TJ, Dunmur DA, Stegemeyer H. 2004. *Crystals that flow: Classic papers from the history of liquid crystals.* London and New York: Taylor & Francis
10. Takezoe H. 2017. *Molecular Crystals and Liquid Crystals* 646: 46-65

11. Nishikawa H, Shiroshita K, Higuchi H, Okumura Y, Haseba Y, et al. 2017. *Advanced Materials* 29: 1702354
12. Mandle RJ, Cowling SJ, Goodby JW. 2017. *Physical Chemistry Chemical Physics* 19: 11429-35
13. Sebastián N, Cmok L, Mandle RJ, de la Fuente MR, Olenik ID, et al. 2020. *Physical Review Letters* 124: 037801
14. Chen X, Korblova E, Dong DP, Wei XY, Shao RF, et al. 2020. *Proceedings of the National Academy of Sciences of the United States of America* 117: 14021-31
15. Clark NA, Chen X, MacLennan JE, Glaser MA. 2024. *Physical Review Research* 6:013195
16. Adaka A, Rajabi M, Haputhantrige N, Sprunt S, Lavrentovich OD, Jákli A. 2024. *Physical Review Letters* 133: 038101
17. Erkoreka A, Martinez-Perdiguero J. 2024. *Physical Review E* 110: L022701
18. Szydlowska J, Majewski P, Cepic M, Vaupotic N, Rybak P, et al. 2023. *Physical Review Letters* 130: 216802
19. Landau LD, Lifshitz E. 1935. *Phys. Zeitsch. Sowjetunion* 8: 153-69
20. Kittel C. 1946. *Physical Review* 70: 965-71
21. Kelvin L. 1984. *The Molecular Tactics of a Crystal*. Oxford, England: Clarendon Press
22. Planer J. 1861. *Annalen der Chemie und Pharmacie* 118: 25-27
23. Planer J. 2010. *Condensed Matter Physics* 13: 37001
24. Wright DC, Mermin ND. 1989. *Reviews of Modern Physics* 61: 385-432
25. Fall WS, Wensink HH. 2025. *ArXiv*: 2502.16526v1
26. Dzyaloshinsky I. 1958. *Journal of Physics and Chemistry of Solids* 4: 241-55
27. Moriya T. 1960. *Physical Review* 120: 91-98
28. Harris AB, Kamien RD, Lubensky TC. 1999. *Reviews of Modern Physics* 71: 1745-57
29. Borshch V, Kim YK, Xiang J, Gao M, Jákli A, et al. 2013. *Nature Communications* 4: 2635
30. Meyer RB. 1976. In *Molecular Fluids. Les Houches Lectures, 1973*, ed. R Balian, G Weill, pp. 271-343. Les Houches: Gordon and Breach
31. Dozov I. 2001. *Europhysics Letters* 56: 247-53
32. Memmer R. 2002. *Liquid Crystals* 29: 483-96
33. Shamid SM, Dhakal S, Selinger JV. 2013. *Physical Review E* 87: 052503
34. Cestari M, Diez-Berart S, Dunmur DA, Ferrarini A, de la Fuente MR, et al. 2011. *Phys Rev E Stat Nonlin Soft Matter Phys* 84: 031704
35. Chen D, Porada JH, Hooper JB, Klittnick A, Shen YQ, et al. 2013. *Proceedings of the National Academy of Sciences of the United States of America* 110: 15931-36
36. Gao M, Kim YK, Zhang CY, Borshch V, Zhou S, et al. 2014. *Microscopy Research and Technique* 77: 754-72
37. Yu HN, Welch C, Mehl GH. 2025. *Spectrochimica Acta Part a-Molecular and Biomolecular Spectroscopy* 330: 125682
38. Gennes PGd. 1968. *Solid State Communications* 6: 163-65
39. Meyer RB. 1968. *Applied Physics Leters* 12: 281-82
40. Lavrentovich OD. 2020. *Optical Materials Express* 10: 2415-24
41. Iadlovska OS, Thapa K, Rajabi M, Mrukiewicz M, Shiyanovskii SV, Lavrentovich OD. 2024. *MRS Bulletin* 49: 835-50
42. Meyer RB. 1969. *Physical Review Letters* 22: 918-20
43. Prost J, Marcerou JP. 1977. *Journal de Physique* 38: 315-24
44. Lavrentovich OD, Pergamenshchik VM. 1995. *International Journal of Modern Physics B* 9: 2389-437
45. Williams RD. 1986. *Journal of Physics A-Mathematical and General* 19: 3211-22

46. Lavrentovich OD, Sergan VV. 1990. *Nuovo Cimento Della Societa Italiana Di Fisica D-Condensed Matter Atomic Molecular and Chemical Physics Fluids Plasmas Biophysics* 12: 1219-22
47. Prinsen P, van der Schoot P. 2004. *Journal of Physics-Condensed Matter* 16: 8835-50
48. Tortora L, Lavrentovich OD. 2011. *Proceedings of the National Academy of Sciences of the United States of America* 108: 5163-68
49. Vanzo D, Ricci M, Berardi R, Zannoni C. 2012. *Soft Matter* 8: 11790-800
50. Peng CH, Lavrentovich OD. 2015. *Soft Matter* 11: 7257-63
51. Wang XG, Bukusoglu E, Miller DS, Pantoja MAB, Xiang J, et al. 2016. *Advanced Functional Materials* 26: 7343-51
52. Lavrentovich OD, Terent'ev EM. 1986. *Sov. Phys. JETP* 64: 1237-44
53. Candau S, Leroy P, Debeauvais F. 1973. *Molecular Crystals and Liquid Crystals* 23: 283-97
54. Poulin P, Stark H, Lubensky TC, Weitz DA. 1997. *Science* 275: 1770-73
55. Poulin P, Weitz DA. 1998. *Physical Review E* 57: 626-37
56. Stark H. 2001. *Physics Reports-Review Section of Physics Letters* 351: 387-474
57. Omori EK, Masso GHX, Biagio RL, Evangelista LR, de Souza RT, Zola RS. 2023. *Liquid Crystals* 50: 1392-405
58. Cladis PE, Kléman M. 1972. *Journal de Physique* 33: 591-98
59. Nayani K, Chang R, Fu JX, Ellis PW, Fernandez-Nieves A, et al. 2015. *Nature Communications* 6: 8067
60. Jeong J, Kang L, Davidson ZS, Collings PJ, Lubensky TC, Yodh AG. 2015. *Proceedings of the National Academy of Sciences of the United States of America* 112: E1837-E44
61. Dietrich CF, Rudquist P, Lorenz K, Giesselmann F. 2017. *Langmuir* 33: 5852-62
62. Myers L, Vinals J. 2025. *Soft Matter* 21: 3768-81
63. Press MJ, Arrott AS. 1974. *Physical Review Letters* 33: 403-06
64. Lavrentovich OD. 1992. *Physical Review A* 46: R722-R25
65. Yi Y, Clark NA. 2013. *Liquid Crystals* 40: 1736-47
66. Jeong J, Davidson ZS, Collings PJ, Lubensky TC, Yodh AG. 2014. *Proceedings of the National Academy of Sciences of the United States of America* 111: 1742-47
67. Nych A, Ognysta U, Musevic I, Sec D, Ravnik M, Zumer S. 2014. *Physical Review E* 89: 062502
68. Davidson ZS, Kang L, Jeong J, Still T, Collings PJ, et al. 2015. *Physical Review E* 92: 019905
69. Mushenheim PC, Pendery JS, Weibel DB, Spagnolie SE, Abbott NL. 2016. *Proceedings of the National Academy of Sciences of the United States of America* 113: 5564-69
70. Martinez A, Collings PJ, Yodh AG. 2018. *Physical Review Letters* 121: 177801
71. Zhou S, Nastishin YA, Omelchenko MM, Tortora L, Nazarenko VG, et al. 2012. *Physical Review Letters* 109: 037801
72. Zhou S, Neupane K, Nastishin YA, Baldwin AR, Shiyanovskii SV, et al. 2014. *Soft Matter* 10: 6571-81
73. Zhou S, Cervenka AJ, Lavrentovich OD. 2014. *Physical Review E* 90: 042505
74. Dietrich CF, Collings PJ, Sottmann T, Rudquist P, Giesselmann F. 2020. *Proceedings of the National Academy of Sciences of the United States of America* 117: 27238-44
75. Varytimiadou S, Revignas D, Giesselmann F, Ferrarini A. 2024. *Liquid Crystals Reviews* 12: 57-104
76. Lavrentovich OD. 2024. *Liquid Crystals Reviews* 12: 1-13
77. Luk'yanchuk I, Tikhonov Y, Razumnaya A, Vinokur VM. 2020. *Nature Communications* 11: 2433
78. Chen X, Korblova E, Glaser MA, Maclennan JE, Walba DM, Clark NA. 2021. *Proceedings of the National Academy of Sciences of the United States of America* 118: e2104092118
79. Nishikawa H, Araoka F. 2021. *Advanced Materials* 33: 2101305
80. Feng C, Saha R, Korblova E, Walba DM, Sprunt SN, Jákli A. 2021. *Advanced Optical Materials* 9: 2101230

81. Zhao X, Zhou J, Li J, Kougo J, Wan Z, et al. 2021. *Proceedings of the National Academy of Sciences of the United States of America* 118: e2111101118
82. Nazarenko KG, Kasian NA, Minenko SS, Samoilov OM, Nazarenko VG, et al. 2023. *Liquid Crystals* 50: 98-109
83. Araoka F, Nishikawa H. 2024. *Molecular Crystals and Liquid Crystals* 768: 1075-83
84. Thapa K, Iadlovska OI, Basnet B, Wang H, Paul A, et al. 2024. *Physical Review E* 109: 054702
85. Pociecha D, Walker R, Cruickshank E, Szydlowska J, Rybak P, et al. 2022. *Journal of Molecular Liquids* 361: 119532
86. Meyer RB. 1969. *Applied Physics Letters* 14: 208-09
87. Xiang J, Li YN, Li Q, Paterson DA, Storey JMD, et al. 2015. *Advanced Materials* 27: 3014-18
88. Karcz J, Herman J, Rychlowicz N, Kula P, Górecka E, et al. 2024. *Science* 384: 1096-99
89. Nishikawa H, Okada D, Kwaria D, Nihonyanagi A, Kuwayama M, et al. 2024. *Advanced Science* 2024: 202405718
90. Gibb CJ, Hobbs J, Nikolova DI, Raistrick T, Berrow SR, et al. 2024. *Nature Communications* 15: 5845
91. Strachan GJ, Górecka E, Szydlowska J, Makal A, Pociecha D. 2024. *Advanced Science* 2024: 2409754
92. Chen X, Martinez V, Nacke P, Korblova E, Manabe A, et al. 2022. *Proceedings of the National Academy of Sciences of the United States of America* 119: e2210062119
93. Song YH, Deng MH, Wang ZD, Li JX, Lei HY, et al. 2022. *Journal of Physical Chemistry Letters* 13: 9983-90
94. Kikuchi H, Matsukizono H, Iwamatsu K, Endo S, Anan S, Okumura Y. 2022. *Advanced Science* 9: 2202048
95. Matsukizono H, Iwamatsu K, Endo S, Okumura Y, Anan S, Kikuchi H. 2023. *Journal of Materials Chemistry C* 11: 6183-90
96. Hobbs J, Gibb CJ, Mandle RJ. 2024. *Small Science* 4: 2400189
97. Nakasugi S, Kang S, Chang TFM, Manaka T, Ishizaki H, et al. 2023. *Journal of Physical Chemistry B* 127: 6585-95
98. Hobbs J, Gibb CJ, Pociecha D, Szydlowska J, Górecka E, Mandle RJ. 2024. *Angewandte Chemie-International Edition* 2024: e202416545
99. Liao Q, Aya S, Huang M. 2024. *Liquid Crystals Reviews* 12: 149-94
100. Brown S, Cruickshank E, Storey JMD, Imrie CT, Pociecha D, et al. 2021. *Chemphyschem* 22: 2506-10
101. Chen X, Martinez V, Korblova E, Freychet G, Zhernenkov M, et al. 2023. *Proceedings of the National Academy of Sciences of the United States of America* 120: e2217150120
102. Sebastián N, Čopič M, Mertelj A. 2022. *Physical Review E* 106: 021001
103. Nishikawa H, Okumura Y, Kwaria D, Nihonyanagi A, Araoka F. 2025. *Adv. Mater.* 37(26):2501946
104. Erkoreka A, Mertelj A, Huang M, Aya S, Sebastián N, Martinez-Perdiguero J. 2023. *Journal of Chemical Physics* 159: 184502
105. Nacke P, Tuffin R, Klasen-Memmer M, Rudquist P, Giesselmann F. 2024. *Scientific Reports* 14: 15018
106. Thoen J, Cordoyiannis G, Korblova E, Walba DM, Clark NA, et al. 2024. *Physical Review E* 110: 014703
107. Rupnik PM, Hanzel E, Lovsin M, Osterman N, Gibb CJ, et al. 2025. *Advanced Science* 12: 2414818
108. Song YH, Li JX, Xia RL, Xu H, Zhang XX, et al. 2022. *Physical Chemistry Chemical Physics* 24: 11536-43
109. Karcz J, Rychlowicz N, Czarnecka M, Kocot A, Herman J, Kula P. 2023. *Chemical Communications* 59: 14807-10
110. Cruickshank E, Rybak P, Majewska MM, Ramsay S, Wang C, et al. 2023. *ACS Omega* 8: 36562-68

111. Nacke P, Manabe A, Klasen-Memmer M, Chen X, Martinez V, et al. 2024. *Scientific Reports* 14: 4473
112. Mertelj A, Cmok L, Sebastián N, Mandle RJ, Parker RR, et al. 2018. *Physical Review X* 8: 041025
113. Rosseto MP, Selinger JV. 2020. *Physical Review E* 101: 052707
114. Čopič M, Mertelj A. 2020. *Physical Review E* 101: 022704
115. Dhakal S, Selinger JV. 2010. *Physical Review E* 81: 031704
116. Zhuang Z, Maclennan JE, Clark NA. 1989. *Liquid Crystal Chemistry, Physics, and Applications* 1080: 110-14
117. Ma Z, Jiang M, Sun A, Yi S, Yang J, et al. 2025. *Phys. Rev. Lett.* 134:238101
118. Ghimire A, Basnet B, Wang H, Guragain P, Baldwin A, et al. 2025. *Soft Matter* 21:8510-22
119. Caimi F, Nava G, Fuschetto S, Lucchetti L, Paie P, et al. 2023. *Nature Physics* 19: 1658-66
120. Marchenko AA, Kapitanchuk OL, Lopatina YY, Nazarenko KG, Senenko AI, et al. 2024. *Physical Review Letters* 132: 098101
121. Caimi F, Nava G, Barboza R, Clark NA, Korblova E, et al. 2021. *Soft Matter* 17: 8130-39
122. Basnet B, Rajabi M, Wang H, Kumari P, Thapa K, et al. 2022. *Nature Communications* 13: 3932
123. Yu JS, Lee JH, Lee JY, Kim JH. 2023. *Soft Matter* 19: 2446-53
124. Basnet B, Paladugu S, Kurochkin O, Buluy O, Aryasova N, et al. 2025. *Nature Communications* 16: 1444
125. Ramdane OO, Auroy P, Forget S, Raspaud E, Martinot-Lagarde P, Dozov I. 2000. *Physical Review Letters* 84: 3871-74
126. Kumari P, Basnet B, Wang H, Lavrentovich OD. 2023. *Nature Communications* 14: 748
127. Manabe A, Bremer M, Kraska M. 2021. *Liquid Crystals* 48: 1079-86
128. Li JX, Xia RL, Xu H, Yang JD, Zhang XX, et al. 2021. *Journal of the American Chemical Society* 143: 17857-61
129. Li JX, Wang ZD, Deng MH, Zhu YY, Zhang XX, et al. 2022. *Giant* 11: 100109
130. Yang JD, Zou Y, Tang WT, Li JX, Huang MJ, Aya S. 2022. *Nature Communications* 13: 7806
131. Perera K, Saha R, Nepal P, Dharmarathna R, Hossain MS, et al. 2023. *Soft Matter* 19: 347-54
132. Meyer RB, Liebert L, Strzelecki L, Keller P. 1975. *Journal De Physique Lettres* 36: L69-L71
133. Lavrentovich MO, Kumari P, Lavrentovich OD. 2025. *Nature Communications* 13: 3932
134. Meyer RB. 1977. *Molecular Crystals and Liquid Crystals* 40: 33-48
135. Okano K. 1986. *Japanese Journal of Applied Physics Part 2-Letters* 25: L846-L47
136. Lee JB, Pelcovits RA, Meyer RB. 2007. *Physical Review E* 75: 051701
137. Ackerman PJ, Smalyukh II. 2017. *Nature Materials* 16: 426-32
138. Ackerman PJ, Smalyukh II. 2017. *Physical Review X* 7: 011006
139. Smalyukh II. 2020. *Reports on Progress in Physics* 83: 106601
140. Wu JS, Smalyukh II. 2022. *Liquid Crystals Reviews* 10: 34-68
141. Fukuda J. 2022. *Liquid Crystals Reviews* 10: 69-90
142. Kumari P, Kurochkin O, Nazarenko VG, Lavrentovich OD, Golovaty D, Sternberg P. 2024. *Physical Review Research* 6: 043207
143. Hedlund KG, Martinez V, Chen X, Park CS, Maclennan JE, et al. 2025. *Physical Chemistry Chemical Physics* 27: 119-28
144. Pattanaporkratana A. 2008. *Textures and interactions between vortices in the 2D XY field of freely suspended SmC and SmC\* liquid crystal films, PhD thesis.* University of Colorado, Boulder, CO. 133 pp.
145. Friedel G, Grandjean F. 1910. *Comptes rendus de l-Acadèmie des Sciences* 151: 762-65
146. Rosenblatt CS, Pindak R, Clark NA, Meyer RB. 1977. *Journal de Physique* 38: 1105-15
147. Kumari P, Basnet B, Lavrentovich MO, Lavrentovich OD. 2024. *Science* 383: 1364-68
148. Grönfors E, Rudquist P. 2025. *Liquid Crystals*: 10.1080/02678292.2025.2470124

149. Khachaturyan AG. 1975. *Journal of Physics and Chemistry of Solids* 36: 1055-61
150. Paik L, Selinger JV. 2025. *Phys. Rev. E* 111:Lo53402
151. Yi S, Hong Z, Ma Z, Zhou C, Jiang M, et al. 2024. *PNAS* 121: e2413879121
152. Sebastián N, Mandle RJ, Petelin A, Eremin A, Mertelj A. 2021. *Liquid Crystals* 48: 2055-71
153. Lavrentovich OD, Nastishin YA. 1990. *Europhysics Letters* 12: 135-41
154. Tran L, Lavrentovich MO, Beller DA, Li NW, Stebe KJ, Kamien RD. 2016. *Proceedings of the National Academy of Sciences of the United States of America* 113: 7106-11
155. Lonberg F, Meyer RB. 1985. *Physical Review Letters* 55: 718-21
156. Srajer G, Lonberg F, Meyer RB. 1991. *Physical Review Letters* 67: 1102-05
157. Chen X, Pecinovsky C, Korblova E, Glaser MA, Radzihovsky L, et al. 2024. *ArXiv*: 2412.19061
158. Kléman M, Michel L, Toulouse G. 1977. *Journal De Physique Lettres* 38: L195-L97
159. Volovik GE, Mineyev VP. 1977. *Sov. Phys. JETP* 45: 1186-96
160. Volovik GE, Lavrentovich OD. 1983. *Sov. Phys. JETP* 58:1159-66
161. Lavrentovich OD. 2020. *Proceedings of the National Academy of Sciences of the United States of America* 117: 14629-31
162. Kats EI. 2021. *Journal of Experimental and Theoretical Physics* 132: 641-44
163. Volovik GE, Zhang K. 2020. *Physical Review Research* 2: 023263
164. Pocieha D, Szydlowska J, et al. 2025. *Adv.Sci.* 2025: e08405